\documentclass[12pt,preprint]{aastex}

\newcommand{\msun}{M_{\odot}}

\shorttitle{The Homogeneity of Interstellar Krypton}
\shortauthors{Cartledge, Meyer, \& Lauroesch}

\begin{document}


\title{The Homogeneity of Interstellar Krypton in the Galactic
Disk\footnote{Based on observations with the NASA/ESA \textit{Hubble Space
Telescope (HST)} and the NASA-CNES-CSA \textit{Far-Ultraviolet Spectroscopic
Explorer (FUSE)}. \textit{HST} spectra were obtained at the Space Telescope
Science Institute, which is operated by the Association of Universities for
Research in Astronomy, Inc. under NASA contract NAS 5-26555; \textit{FUSE} is
operated for NASA by the Johns Hopkins University under NASA contract
NAS-32985.}}


\author{Stefan I. B. Cartledge}
\affil{Department of Physics and Astronomy, Louisiana State University,
    Baton Rouge, LA 70803}
\email{scartled@lsu.edu}

\and

\author{David M. Meyer and J. T. Lauroesch}
\affil{Department of Physics and Astronomy, Northwestern University,
    Evanston, IL 60208}
\email{davemeyer@northwestern.edu, jtl@elvis.astro.nwu.edu}





\begin{abstract}
We present an analysis of high resolution \textit{HST} Space Telescope Imaging
Spectrograph (STIS) observations of \ion{Kr}{1} $\lambda$1236 absorption in
seven sight lines that probe a variety of interstellar environments. In
combination with krypton and hydrogen column densities derived from current and
archival STIS and \textit{Far-Ultraviolet Spectroscopic Explorer} data, the
number of sight lines with reliable {Kr/H} ISM abundance ratios has been
increased by 50\% to 26---including paths that sample a range of nearly 5
orders of magnitude in $f$(H$_2$), over 2 orders of magnitude in
${\langle}n_{\rm H}\rangle$, and extending up to 4.8 kpc in length. For sight
lines contained entirely within the local spiral arm (the Orion Spur), the
spread of Kr/H ratios about the mean of log$_{10}$[N(Kr)/N(H)]$_{ISM}$ =
$-9.02\pm0.02$ is remarkably tight (0.06 dex), less than the typical datapoint
uncertainty. Intriguingly, the only two sight lines that extend through
neighboring structures, in particular gas associated with the
Carina/Sagittarius Arm, exhibit relatively large, near-solar krypton abundances
(log$_{10}$[N(Kr)/N(H)]$_{combined}$ = $-8.75^{+0.09}_{-0.11}$). Although these
deviations are only measured at the 2$\sigma$ level, they suggest the
possibility that krypton abundances beyond the Orion Spur may differ from the
local value.
\end{abstract}




\section{Introduction}
The detection and measurement of absorption features engendered by interstellar
krypton became possible with the launch of the \textit{Hubble Space Telescope
(HST)} and the Goddard High Resolution Spectrograph (GHRS) in 1990; since that
time krypton has garnered significant interest as a direct probe of elemental
abundances in the interstellar medium (ISM) and, consequently, of the nature
and magnitude of the effects of various interstellar processes. Several factors
make krypton particularly useful in this regard. Since krypton is a noble
gas with a symmetric electron configuration, essentially all interstellar
krypton should be found in the gas phase. Furthermore, since its ionization
potential is larger than that of hydrogen, one need only examine resonant
absorption features for the neutral form to determine the krypton abundance
along a particular sight line. Finally, the intrinsic strengths of the atomic
lines and krypton's low relative abundance to hydrogen combine to place them on
the linear portion of the curve of growth for interstellar paths with a wide
variety of absorption properties, enabling the accurate measurement of krypton
abundances by direct examination of UV absorption features.

GHRS achieved the first reliable detections of interstellar krypton using its
unique combination of improved UV sensitivity and spectral resolution with
respect to previous instruments \citep{car91}. Consequently, \citet{car97} were
able to show that 10 GHRS-observed sight lines probing the local diffuse ISM
within about 500 pc possessed nearly identical Kr/H abundance ratios, consistent
with the expectation that interstellar krypton is undepleted. \citet{car01}
broadened the scope of krypton study by presenting Kr/H measurements for seven
sight lines observed with both the Space Telescope Imaging Spectrograph (STIS)
and the \textit{Far Ultraviolet Spectroscopic Explorer (FUSE)}, including five
that intersect translucent clouds.
Although these sight lines provided some evidence for enhanced oxygen depletion,
no krypton abundance variations distinguishable from measurement uncertainty
were apparent, implying a consistent Kr/H ratio for both the diffuse and
translucent ISM. The number of sight lines that have been studied to date,
however, is still small and samples a limited physical space. In this Letter,
we supplement previous krypton abundances derived from GHRS and STIS data with
krypton results for seven new sight lines and hydrogen measurements based on
STIS and \textit{FUSE} data for nine, increasing the total number with reliable
krypton and hydrogen column densities by about 50\% to 26. The new sight lines
include the first to intersect krypton gas associated with a neighboring
spiral arm; intriguingly, these sight lines are the only ones to exhibit
near-solar Kr/H abundance ratios.

\section{Observations and Measurements}

The current STIS observations of \ion{Kr}{1} $\lambda$1236 absorption toward
seven Galactic O and B stars were garnered through a single \textit{HST} Cycle
8 observing program (GO8241) running from July 1999 until April 2001; a few
early results derived from this program were included in the \citet{car01}
sample. All new krypton spectra were acquired using the E140H echelle grating
and {0.2\arcsec$\times$0.2\arcsec} STIS aperture with a 202{\AA} spectral
window centered at 1271{\AA}. Initial processing of the raw data was
accomplished using the standard STSDAS STIS data reduction package to produce
geometrically corrected two dimensional flat fielded spectra. In the interest
of consistency with \citet{car01}, the \citet{how00} scattered-light correction
algorithm was then applied to generate one dimensional extracted spectra. In
general, \ion{Kr}{1} $\lambda$1236 absorption appeared in the overlap region
between successive spectral orders, which were combined to improve the S/N
ratio. The final S/N per pixel values near 1235{\AA} ranged from 25--40.

The determination of column densities based on \ion{Kr}{1} $\lambda$1236
absorption profiles proceeded according to two distinct methods in order to
evaluate and correct for uncertainties due to line saturation: profile fitting
\citep{mar95,wel91} and measurement of apparent optical depth as a function of
wavelength \citep{sav91}. Each method was applied in the same manner and with
the same physical constants used by \citet{car01}; the current results of both
profile fitting and apparent optical depth methods are presented in
Table~\ref{kryptab}. Notably, the new column densities derived by each method
differ by less than 0.03 dex for an individual sight line, and these results
are only up to 0.06 dex larger than values derived under the assumption of no
saturation. Since individual measurement uncertainties are of order 0.05--0.11
dex, it is not expected that unresolved saturation is a serious concern for
these new sight lines. The results of profile-fitting measurements are adopted
for all further analysis and are listed in Table~\ref{kryhydtab}, which also
summarizes previous GHRS and STIS measurements.

Hydrogen column densities were derived in all cases where Ly-$\alpha$ absorption
in the STIS data appeared to be uncontaminated by a stellar contribution and
where \textit{FUSE} data existed to provide access to H$_2$. STIS spectral
orders covering the 1160--1280{\AA} wavelength interval were averaged together
to construct an atomic hydrogen absorption profile for each sight line and the
associated column density was determined using the continuum-reconstruction
method \citep{boh75,dip94}. Notably, four sight lines with newly-measured
hydrogen abundances were examined by \citet{dip94}---the current \ion{H}{1}
value agrees with the previous result to within 0.04 dex in each case.

The continuum-reconstruction technique also proved useful in measuring the
molecular hydrogen column densities appropriate to several rovibrational
absorption profiles apparent in the \textit{FUSE} spectra between 1040 and
1120{\AA}, since the amount of material present in each sight line was
sufficient to generate noticeable damping wings. The \textit{FUSE} LWRS
aperture data used to determine molecular hydrogen abundances were collected
from several programs examining the ISM, including P101, P116, P235, B030, B071,
and Z901. All spectra were processed through CALFUSE, the majority with version
1.8.7. The \ion{H}{1}
and H$_2$ values for all 26 sight lines with reliable Kr/H abundance ratios
appear in Table~\ref{kryhydtab}; however, it should be noted that one of the
seven new krypton sight lines, the path toward HD148594, does not appear in the
table. The HD148594 STIS spectrum implies an unreasonably large atomic hydrogen
column density, likely due to stellar contamination of the Ly-$\alpha$ profile
(its spectral type is B8V). The nine new hydrogen column density determinations
include the remaining six new sight lines and three that were studied by
\citet{car01} (HD37903, HD152590, and HD203532), that have subsequently been
observed by \textit{FUSE}.

%
\section{Local ISM Sight Lines}
\label{section_local_ISM}
\begin{figure}
\plotone{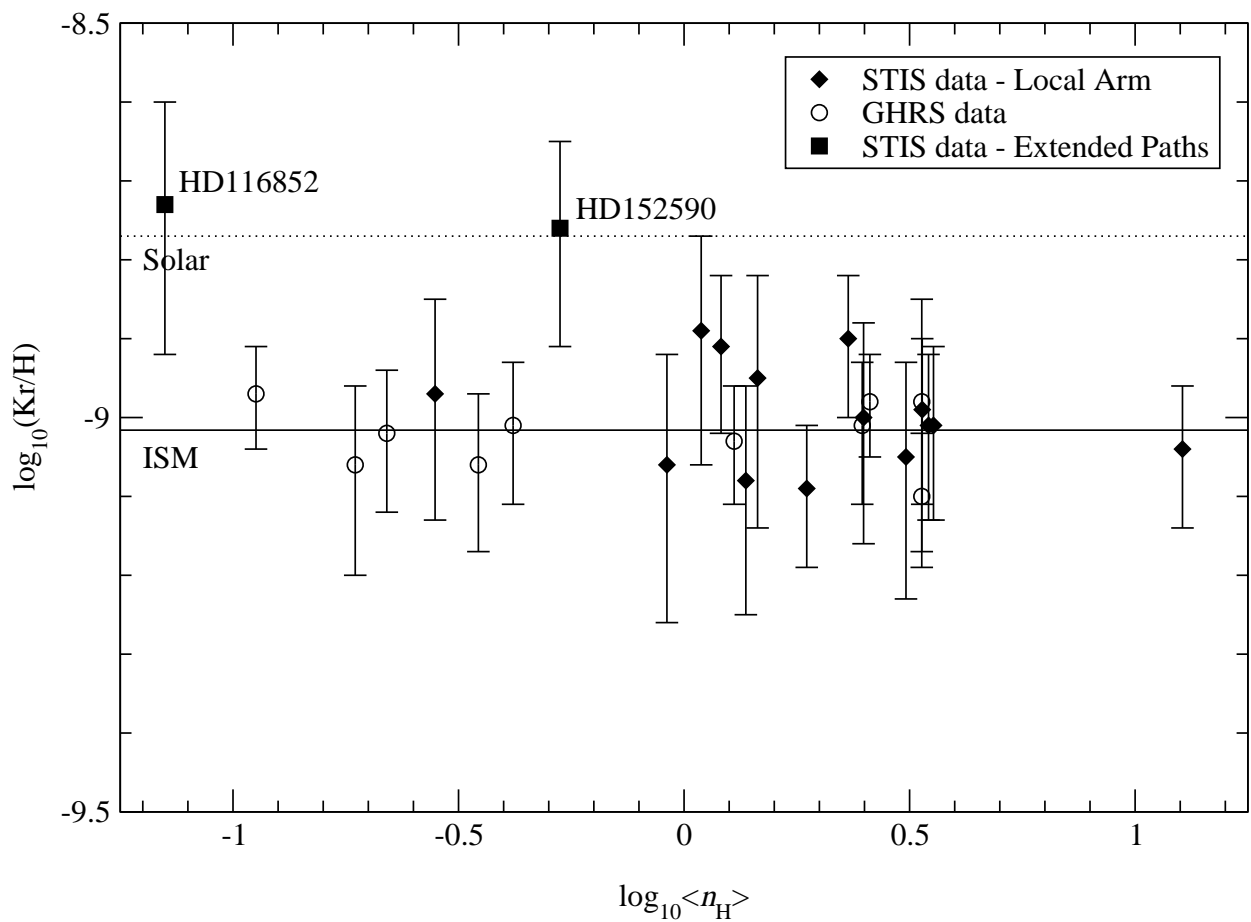}
\caption{Kr/H Abundance Ratios as a Function of Mean Hydrogen Sight Line
Density\newline Krypton abundances are plotted above as a function of the
sight line property ${\langle}n_{\rm H}\rangle$, which, among the properties
listed in Table~\ref{kryhydtab}, distinguishes each path
most clearly and highlights the fact that there is near-unanimous agreement
between each sight line and a single Kr/H within 1$\sigma$ error bars. The only
serious detractors from this accord are paths extending through not just local
gas, but the Carina/Sagittarius Arm ISM as well.\label{fig1}}
\end{figure}

\begin{figure}
\plotone{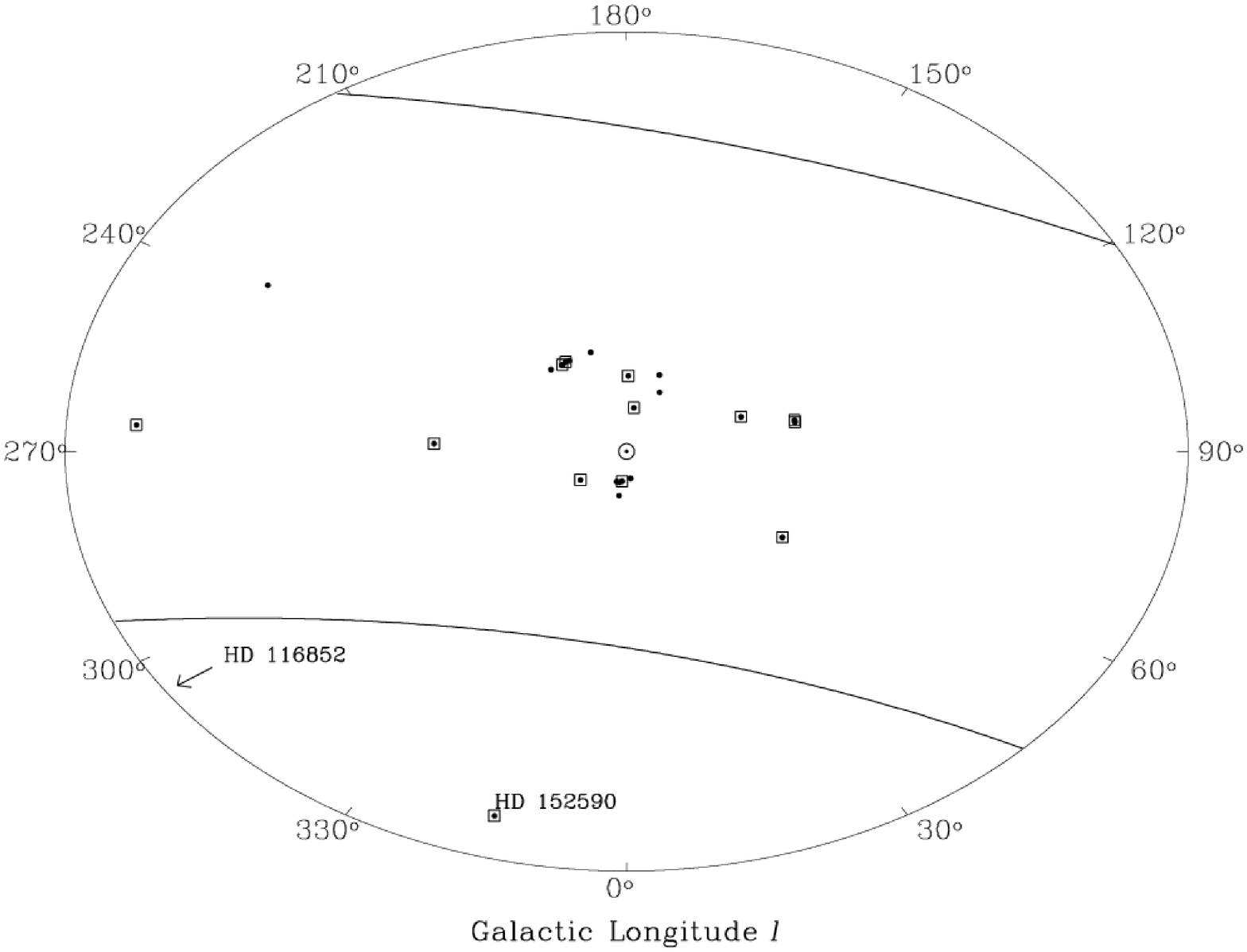}
\caption{STIS and GHRS Sight Lines with Krypton and Hydrogen Abundances\newline
The positions of stars toward which krypton and hydrogen column densities have
been measured by STIS or GHRS and \textit{FUSE} or \textit{Copernicus},
respectively, have been projected onto the Galactic Plane and plotted above in
terms of $l$ and distance; stars observed by STIS are identified by squares. A
variety of distances and directions within the Galactic disk are probed by the
full data set, yet the only sight lines significantly at odds with a single
Kr/H value, those toward HD116852 and HD152590, extend beyond the
Carina/Sagittarius Arm. The solid arcs indicate positions for the
Carina/Sagittarius (center-ward) and Perseus (anti-center) Arms as described by
\citet{val02}; the plot encompasses the Galactic disk within 2 kpc.\label{fig2}}
\end{figure}

A striking feature of the tabulated Kr/H abundance ratios is the uniformity of
values for sight lines that sample interstellar gas associated only with the
local spiral arm, the Orion Spur. This accord is clearly shown in
Figure~\ref{fig1}, in which the Kr/H abundance ratio is plotted as a function
of mean total hydrogen sight line density ${\langle}n_{\rm H}\rangle$. Although
the data set includes sight lines which sample a diverse population of
interstellar environments, as indicated by the wide range in properties such as
mean density, molecular hydrogen fraction, color excess, direction, and
distance (see Table~\ref{kryhydtab} and Figure~\ref{fig2}), each datapoint
associated with the local arm is consistent with the weighted ISM mean
log$_{10}$(Kr/H) = $-9.02\pm0.02$. In fact, the spread among sight lines
confined to sampling Orion Spur gas, as indicated by the standard deviation of
their Kr/H abundance ratios (0.06 dex), is smaller than the typical measurement
uncertainty (0.08--0.10 dex). This result is notable, since the observed
distribution must include contributions from both intrinsic scatter and
measurement error sources. Because these factors are uncorrelated, one must
conclude that although the reported errors appear to be somewhat conservative
estimates, the tightness of this distribution implies that the homogeneity of
the ISM is very narrowly constrained on length scales of a few hundred parsecs.
In particular, the observed variation of the interstellar krypton abundance in
the Orion Spur can be explained by measurement uncertainty alone.

Similarly uniform gas-phase abundance ratios have been noted for GHRS
observations of carbon, nitrogen, and oxygen \citep{car96,mey97,sof97,mey98} in
sight lines sampling a range of environments similar to those probed by the
krypton sight lines of \citet{car97}. In fact, the abundance ratio scatter in
each of the GHRS data sets for these elements and krypton was about 0.06
dex---a surprising result considering that carbon and oxygen are the two
elements most abundant in dust while nitrogen and krypton are essentially
undepleted in the diffuse ISM \citep{sof01}. The addition of several STIS
krypton abundance measurements, including values for both longer and denser
sight lines, makes a more robust case for the homogeneity of the local ISM,
extending the apparent region of its prevalence to several hundred parsecs near
the Sun.

One must bear certain caveats in mind, however, when interpreting the
significance of the local ISM Kr/H gas-phase abundance ratio uniformity. For
instance, the column density quoted for each sight line is integrated along its
entire length and thus the quantity Kr/H is a ratio of integrated values. As a
result, any low amplitude departures from the mean on relatively small spatial
scales are not measurable. But although the degree of increase is in dispute,
analytical and numerical models of ISM mixing processes agree that their
timescales increase with lengthscale \citep{roy95,dea02}. Hence, the close
agreement of hundred-parsec--scale Kr/H ratios indicates that the ISM is
generally well-mixed to at least the level of measurement uncertainty, without
excluding the possibility of significant small-scale departures (e.g.,
enrichment). This conclusion is particularly relevant to the two sight lines
whose Kr/H abundance ratios diverge from the Orion Spur ISM mean since, by
their coincident values, they suggest that the krypton abundances in other
spiral arms might be different from the local value.

\section{HD116852 and HD152590: Elevated Krypton Abundances?}
\label{section_hd116852_hd152590}
The sight lines toward HD116852 and HD152590, in addition to each possessing
a near-solar Kr/H gas-phase abundance ratio (log$_{10}$(Kr/H)$_\odot$ =
$-8.77\pm0.07$; \citealt{and89}), share the distinction of extending
through gas associated with the Carina/Sagittarius Arm. However, these two paths
are widely separated in space from each other and from the bulk of the data set
sight lines (see Figure~\ref{fig2}). \citet{sem94} and \citet{sem96} examined
GHRS spectra of HD116852, placing the star at 4.8 kpc (1.3 kpc below the
Galactic mid-plane) and identifying eight components comprising three large
gas complexes which they associated with local gas and material co-rotating with
the Carina/Sagittarius and Norma Arms. The GHRS data contained absorption
profiles for several elements, but krypton was not included. Final S/N ratios
for the current STIS data do not approach the values achieved using GHRS,
limiting reliable measurement of absorption to the three strongest components.
In accord with the \citet{sem94} summary of the sight line, these absorption
features are a blend associated with gas in the Orion Spur and the
Carina/Sagittarius Arm.

The study of visual absorption in the HD152590 direction has indicated that
this sight line intersects two broad regions of gas spatially coincident with
local gas and the Carina/Sagittarius Arm \citep{rab97}. Although the krypton
column density for this sight line was previously published by \citet{car01},
the corresponding hydrogen value was absent. Consequently, it was not possible
to distinguish between a large krypton or low oxygen abundance to explain the
sight line's unusual O/Kr gas-phase abundance ratio. Since the first STIS visit
to this target, HD152590 has also been observed as a target of the \textit{HST}
SNAPSHOT program GO9434. The second spectrum, however, was acquired using a
lower resolution setup (the E140M grating centered at 1735{\AA} with $\Delta v
\approx$ 6.5 km s$^{-1}$). The poorer resolution of this data, coupled with its
shorter exposure time and a flex in the continuum near the krypton feature,
reduced the quality of krypton abundance measurement significantly, producing
the result log$_{10}N$(Kr)$_{\rm HD152590}$ = $12.61\pm0.11$. This value is
lower than but still consistent with the number in Table~\ref{kryhydtab}, and
although the complications already noted would likely reduce the apparent
krypton abundance, it is possible that the abundance ratio for the HD152590
sight line is less extreme than it currently seems. It should also be noted
that since the abundances presented here are integrated values, any apparent
gaps between local and distant-spiral-arm Kr/H ratios have been mitigated by
local gas in both the HD152590 and HD116852 sight lines.


The krypton enhancements apparent toward HD116852 and HD152590 exist only at
the 2$\sigma$ level, but their deviation from the local ISM mean is highlighted
by the very tight agreement of Orion Spur sight lines and the coincidence that
these two are the only paths to intersect another spiral arm. If they are
considered along with the bulk of the sample, the scatter rises only to about
the level of datapoint uncertainty and the concept of an homogeneous ISM might
conceivably be extended to bridge the inter-arm gaps. However, the agreement
between these Kr/H ratios suggests the possibility that the krypton abundance
beyond the Orion Spur, in particular in the Carina/Sagittarius Arm, may be
measurably different than it is locally. The relevance of such a possibility is
demonstrated by the existence of the galactocentric radial abundance gradient
($-0.07$ dex kpc$^{-1}$ in O/H), that has been identified using stellar
atmosphere, \ion{H}{2} region, and planetary nebula measurements
\citep{hen99,rol00}. The gradient itself, however, is of insufficient magnitude
to explain the two atypical krypton ratios. An alternative, that extra krypton
could be liberated from dust to raise the gas-phase abundance is belied by the
accumulated evidence that krypton is undepleted in the ISM and the lack of any
enhancement for strongly depleted elements in these directions \citep{car3a}.
Consequently, if these elevated Kr/H ratios are real, it seems that they must
be nucleosynthetic in origin.

\begin{figure}
\plotone{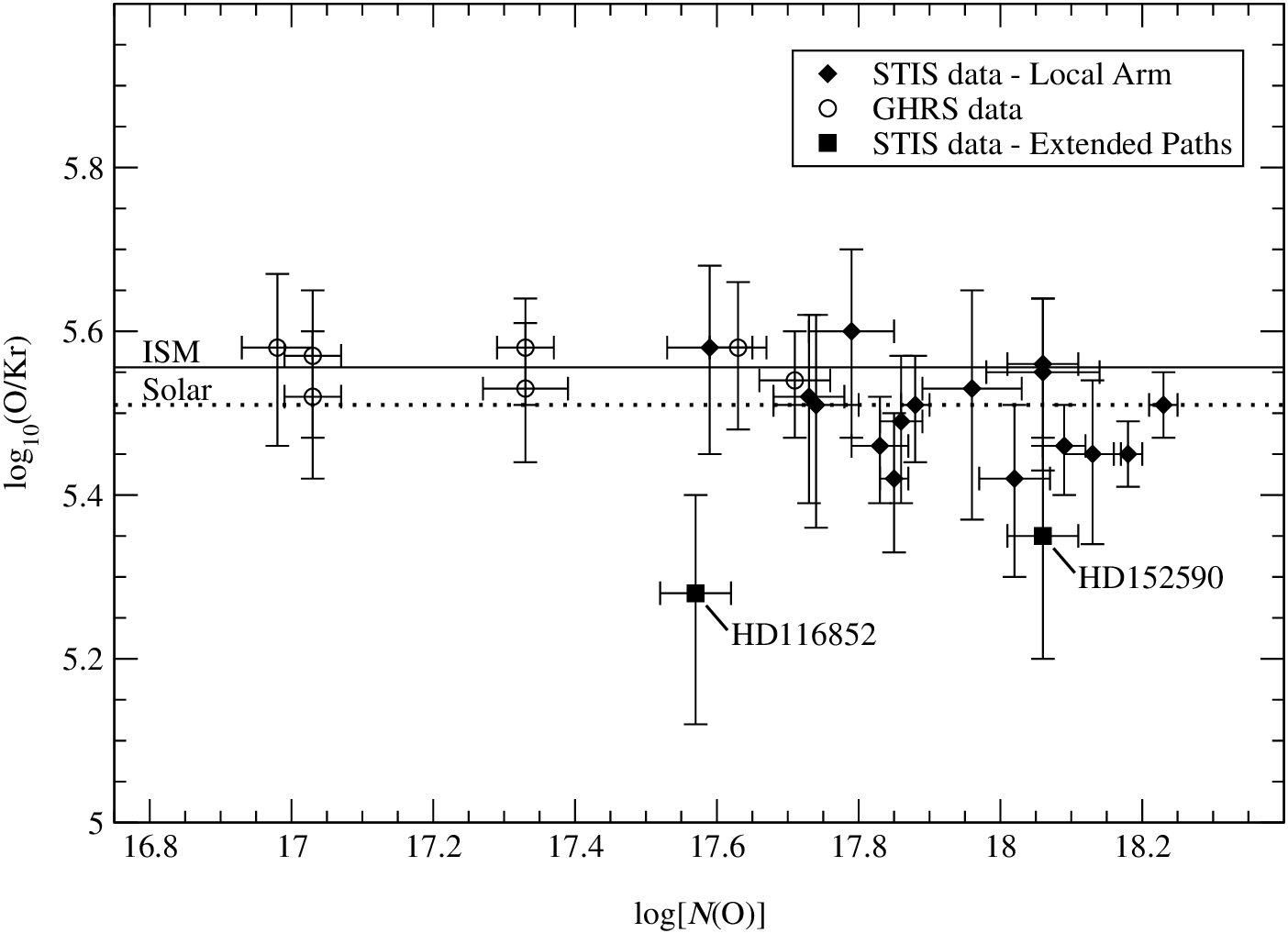}
\caption{O/Kr Abundance Ratios as a Function of Oxygen Column Density\newline
Oxygen--to--krypton abundance ratios are plotted above for sight lines probed
using STIS or GHRS; oxygen data are listed in \citet{car3b} and the solar
(dotted line) and ISM (solid line) reference levels are derived from
\citet{and89}, \citet{car97}, \citet{mey98}, and \citet{hol00}. All sight lines
concur with the ISM limit except the HD116852 and HD152590 paths and several
with large values of ${\langle}n_{\rm H}\rangle$ that appear to exhibit
enhanced oxygen depletion; these datapoints all lie below the line.\label{fig3}}
\end{figure}

In this scenario, a potential source can be identified by comparing krypton's
production sites with those of oxygen. As shown in Figure~\ref{fig3} (from
\citealt{car3b}), these elements' abundances are tightly linked. The gas-phase
O/Kr ratio departs most dramatically from the ISM mean for the sight lines
toward HD116852 and HD152590. Oxygen is predominantly produced in massive star
evolution, which also contributes the majority of the krypton nucleosynthesis
\citep{and89}. However, a large fraction of the interstellar krypton abundance
is derived from the main $s$-process operating in low-mass stars (1--3 $\msun$)
evolving along the asymptotic giant branch \citep{rai93}. Consequently, these
objects are prime candidates for explaining any krypton enhancement toward
HD116852 and HD152590. Notably, this source has recently been cited for the
enhancement of other $s$-process elemental abundances. \citet{sof99} examined
the interstellar abundances of tin and cadmium and attributed well-determined
solar and even supersolar tin abundances to the influence of enrichment by
low--to--intermediate mass stars. STIS spectra covering the 1400.44{\AA}
\ion{Sn}{2} line exist for both HD116852 and HD152590, through GO8662 and
GO9434 observations, respectively. Unfortunately, the latter spectrum is not of
sufficient resolution or S/N to reliably fix the continuum and produce a
useful measurement. However, the gas-phase tin abundance toward HD116852
\emph{can} be determined and it is approximately solar (log$_{10}$(Sn/H)$_{\rm
HD116852}$ = $-9.80\pm0.14$; log$_{10}$(Sn/H)$_\odot$ = -9.86, \citealt{gre96}).
Given its large uncertainty and the possibility that some of the tin in this
sight line is depleted from the gas-phase, the HD116852 tin measurement is
consistent with the conjecture of main $s$-process enrichment.

\section{Concluding Remarks}
\label{summary}
The new krypton abundance measurements presented in this paper supplement
earlier results by more fully sampling gas within the Orion Spur and include
data from the first sight lines to examine krypton abundances in a neighboring
spiral arm. As a result, it has been demonstrated that Kr/H abundance ratios in
the local spiral arm ISM are distributed rather tightly about the value
log$_{10}$(Kr/H) = $-9.02\pm0.02$. The improved coverage of the current data
serves to reinforce the conclusion, initially based on previous smaller samples
of GHRS krypton, carbon, oxygen, and nitrogen abundance measurements, that the
local ISM is very well mixed. Nevertheless, the unusual Kr/H abundance ratios
for the sight lines toward HD116852 and HD152590 demonstrate the possibility
that small-scale departures exist and are detectable. In fact, the similarity
in the krypton abundances for these two cases and their common property of
probing gas outside the local spiral arm suggest that a spatial limit to the
efficient mixing of interstellar material may coincide with the boundaries of
the Orion Spur. Testing of this speculation will require the acquisition and
examination of several additional krypton sight lines with long pathlengths.

In closing, we would like to thank the anonymous referee for his/her comments,
and acknowledge the support for this work by STScI through a grant to
Northwestern University.



\acknowledgments

\clearpage

\clearpage

\clearpage

\clearpage

\begin{deluxetable}{lccr@{$\;\!$}lr@{$\;\!$}lr@{$\;\!$}lr@{$\;\!$}l}
\rotate
\tablecaption{New Krypton Measurements \label{kryptab}}
\tablewidth{0pt}
\tablehead{
&\colhead{Heliocentric Velocities}&\colhead{$b$-values}&
\multicolumn{2}{c}{$W_{\lambda1235}$}&\multicolumn{2}{c}{log$_{10}N$(Kr)$_A$}&
\multicolumn{2}{c}{log$_{10}N$(Kr)$_P$}&\multicolumn{2}{c}{log$_{10}N$(Kr)}\\
\colhead{Star}&\colhead{(km s$^{-1}$)}&\colhead{(km s$^{-1}$)}&
\multicolumn{2}{c}{(m\AA)}&\multicolumn{2}{c}{(cm$^{-2}$)}&
\multicolumn{2}{c}{(cm$^{-2}$)}& 
\multicolumn{2}{c}{(cm$^{-2}$)}\\
}
\startdata
HD37367  & 11.5, 16.5                       & 2.3, 2.7                &
7.81&(0.74) & 12.48&(0.04) & 12.51&(0.04) & 12.51&(0.05) \\
HD72754  & 17.1, 20.8, 26.3                 & 1.4, 1.5, 2.7           &
4.40&(0.79) & 12.23&(0.07) & 12.23&(0.07) & 12.23&(0.10) \\
HD116852 & 3.4, 9.7, 15.4                   & 2.1, 1.8, 1.2           &
5.06&(0.95) & 12.29&(0.08) & 12.29&(0.08) & 12.29&(0.11) \\
HD148594 & $-$10.5, $-$4.7                      & 1.6, 2.2                &
5.73&(0.86) & 12.36&(0.06) & 12.37&(0.05) & 12.37&(0.08) \\
HD198478 & $-$21.6, $-$17.3, $-$12.1, $-$7.4        & 2.2, 2.2, 2.0, 1.8      &
9.80&(1.12) & 12.59&(0.05) & 12.60&(0.07) & 12.60&(0.08) \\
HD208440 & $-$29.1, $-$23.1, $-$17.8, $-$12.8, $-$6.3 & 1.3, 2.5, 1.7, 2.8, 1.4 &
7.49&(1.58) & 12.45&(0.09) & 12.43&(0.04) & 12.43&(0.10) \\
HD220057 & $-$16.7, $-$10.6, $-$6.3               & 2.6, 3.0, 3.0           &
3.93&(0.64) & 12.18&(0.07) & 12.19&(0.05) & 12.19&(0.08) \\
\enddata

\tablecomments{The heliocentric velocities and $b$-values listed above specify
the cloud component models used to fit the absorption profiles for each sight
line. The latter three columns refer to krypton column densities determined by
apparent optical depth analysis (subscript $A$) and profile fitting (subscript
$P$), followed by the adopted result.}

\end{deluxetable}

\clearpage

\begin{deluxetable}{lccr@{$\;\!$}lr@{$\;\!$}lr@{$\;\!$}lr@{$\;\!$}lr@{$\;\!$}lr@{.}lc}
\rotate
\tabletypesize{\footnotesize}
\tablecaption{Krypton and Hydrogen Sight Line Properties\tablenotemark{a} \label{kryhydtab}}
\tablewidth{0pt}
\tablehead{
&\colhead{$E(\bv)$}&\colhead{$d_\ast$}&
\multicolumn{2}{c}{\hspace*{-0.15cm}log$_{10}$[$N$(\ion{H}{1})]}&
\multicolumn{2}{c}{\hspace*{-0.1cm}log$_{10}$[$N$(H$_2$)]}&
\multicolumn{2}{c}{log$_{10}$[$N$(H)]}&
\multicolumn{2}{c}{\hspace*{-0.05cm}log$_{10}$[$N$(Kr)]}&
\multicolumn{2}{c}{ }&
\multicolumn{2}{c}{\hspace*{-0.2cm}log$_{10}{\langle}n_{\rm H}\rangle$}&
\colhead{ }\\
\colhead{Star}&\colhead{(mag)}&\colhead{(kpc)}&
\multicolumn{2}{c}{(cm$^{-2}$)}&
\multicolumn{2}{c}{(cm$^{-2}$)}&
\multicolumn{2}{c}{(cm$^{-2}$)}&
\multicolumn{2}{c}{(cm$^{-2}$)}&
\multicolumn{2}{c}{log$_{10}$[Kr/H]}&
\multicolumn{2}{c}{(cm$^{-3}$)}&
\colhead{log$_{10}f$(H$_2$)}\\
}
\startdata
HD27778        & 0.38 & 0.22 & 21.10&(0.12) & 20.72&(0.08) & 21.36&(0.08) &
12.37&(0.05) & $-$8.99&(0.09) &  0&43 & $-$0.34 \\
HD37021        & 0.54 & 0.50 & 21.68&(0.12) & \multicolumn{2}{c}{...} &
21.68&(0.12) & 12.63&(0.04) & $-$9.05&(0.12) &  0&49 & ... \\
HD37061        & 0.53 & 0.50 & 21.73&(0.09) & \multicolumn{2}{c}{...} &
21.73&(0.09) & 12.72&(0.03) & $-$9.01&(0.09) &  0&45 & ... \\
HD37367         & 0.38 & 0.36 & 21.28&(0.09) & 20.53&(0.09) &
21.41&(0.07) & 12.51&(0.05) & $-$8.90&(0.08) &  0&36 & $-$0.58 \\
HD37903        & 0.35 & 0.50 & 21.16&(0.09) & 20.85&(0.07) &
21.46&(0.06) & 12.37&(0.06) & $-$9.09&(0.08) &  0&32 & $-$0.31 \\
HD72754        & 0.33 & 0.69 & 21.18&(0.12) & 20.35&(0.10) &
21.29&(0.10) & 12.23&(0.10) & $-$9.06&(0.14) & $-$0&04 & $-$0.64 \\
HD75309        & 0.21 & 1.75 & 21.08&(0.09) & 20.20&(0.12) & 21.18&(0.08) &
12.21&(0.09) & $-$8.97&(0.12) & $-$0&55 & $-$0.68 \\
HD116852       & 0.21 & 4.80 & 20.96&(0.09) & 19.79&(0.11) &
21.02&(0.08) & 12.29&(0.11) & $-$8.73&(0.11) & $-$1&15 & $-$0.93 \\
HD147888       & 0.51 & 0.15 & 21.71&(0.09) & 20.57&(0.15) & 21.77&(0.08) &
12.73&(0.03) & $-$9.04&(0.08) &  1&07 & $-$0.90 \\
HD152590       & 0.39 & 1.80 & 21.37&(0.06) & 20.47&(0.07) &
21.47&(0.05) & 12.71&(0.05) & $-$8.76&(0.11) & $-$0&28 & $-$0.70 \\
HD185418       & 0.45 & 0.69 & 21.19&(0.09) & 20.71&(0.12) & 21.41&(0.07) &
12.50&(0.06) & $-$8.91&(0.09) &  0&08 & $-$0.40 \\
HD198478       & 0.54 & 0.79 & 21.32&(0.15) & 20.87&(0.15) &
21.55&(0.11) & 12.50&(0.06) & $-$9.05&(0.12) &  0&16 & $-$0.38 \\
HD203532       & 0.29 & 0.25 & 21.27&(0.09) & 20.64&(0.08) &
21.44&(0.07) & 12.43&(0.08) & $-$9.01&(0.10) &  0&55 & $-$0.50 \\
HD207198       & 0.60 & 0.62 & 21.53&(0.07) & 20.83&(0.10) & 21.68&(0.09) &
12.68&(0.08) & $-$9.00&(0.12) &  0&27 & $-$0.55 \\
HD208440       & 0.29 & 0.62 & 21.23&(0.09) & 20.29&(0.07) &
21.32&(0.08) & 12.43&(0.10) & $-$8.89&(0.12) &  0&04 & $-$0.73 \\
HD220057       & 0.23 & 0.44 & 21.17&(0.09) & 20.28&(0.07) &
21.27&(0.07) & 12.19&(0.08) & $-$9.08&(0.12) &  0&14 & $-$0.59 \\
$\zeta$ Per    & 0.30 & 0.40 & 20.81&(0.04) & 20.67&(0.10) & 21.20&(0.06) &
12.17&(0.03) & $-$9.03&(0.06) &  0&11 & $-$0.23 \\
$\epsilon$ Per & 0.11 & 0.31 & 20.42&(0.06) & 19.53&(0.15) & 20.52&(0.06) &
11.46&(0.07) & $-$9.06&(0.09) & $-$0&46 & $-$0.68 \\
$\lambda$ Ori  & 0.12 & 0.50 & 20.79&(0.08) & 19.11&(0.11) & 20.81&(0.07) &
11.80&(0.05) & $-$9.01&(0.08) & $-$0&38 & $-$1.40 \\
$\epsilon$ Ori & 0.05 & 0.50 & 20.46&(0.07) & 16.57&\hspace*{0.3cm}...    &
20.46&(0.07) & 11.40&(0.08) & $-$9.06&(0.10) & $-$0&73 & $-$3.59 \\
$\kappa$ Ori   & 0.04 & 0.50 & 20.53&(0.04) & 15.68&\hspace*{0.3cm}...    &
20.53&(0.04) & 11.51&(0.07) & $-$9.02&(0.08) & $-$0&66 & $-$4.55 \\
$\tau$ CMa     & 0.17 & 1.51 & 20.71&(0.02) & 15.48&\hspace*{0.3cm}...    &
20.71&(0.04) & 11.75&(0.05) & $-$8.97&(0.06) & $-$0&96 & $-$4.93 \\
1 Sco          & 0.19 & 0.16 & 21.21&(0.06) & 19.23&(0.10) & 21.22&(0.06) &
12.24&(0.12) & $-$8.98&(0.13) &  0&53 & $-$1.69 \\
$\delta$ Sco   & 0.16 & 0.16 & 21.08&(0.06) & 19.41&(0.11) & 21.10&(0.03) &
12.12&(0.05) & $-$8.98&(0.06) &  0&41 & $-$1.38 \\
$\omega^1$ Sco & 0.22 & 0.23 & 21.18&(0.08) & 20.05&(0.06) & 21.24&(0.07) &
12.23&(0.05) & $-$9.01&(0.08) &  0&40 & $-$0.89 \\
$\zeta$ Oph    & 0.32 & 0.14 & 20.71&(0.02) & 20.65&(0.05) & 21.15&(0.03) &
12.05&(0.07) & $-$9.10&(0.08) &  0&53 & $-$0.20 \\
\enddata


\tablenotetext{a}{The sight lines identified by HD numbers were observed with
STIS and \textit{FUSE}--the remainder were observed using GHRS and
\textit{Copernicus}. In addition to new Kr and H column densities for HD37367,
HD72754, HD116852, HD198478, HD208440, and HD220057 and new H results for
HD37903, HD152590, and 203532, values are included for the STIS sight lines
appearing in \citet{car01} and GHRS sight lines compiled by \citet{car97}; the
latter have been linearly adjusted to reflect newer $f$-values.}


\end{deluxetable}





\begin{thebibliography}{}
\bibitem[Afflerbach, Churchwell, \& Werner(1997)]{aff97} Afflerbach, A.,
    Churchwell, E., \& Werner, M.~W. 1997, \apj, 478, 190
\bibitem[Anders \& Grevesse(1989)]{and89} Anders, E., \& Werner, M. W. 1989,
    Geochim. Cosmochim. Acta, 53, 197
\bibitem[Bi\'emont \& Zeippen(1992)]{bie92} Bi\'emont, E., \& Zeippen, C. J.
    1992, \aap, 265, 850
\bibitem[Bohlin(1975)]{boh75} Bohlin, R.~C. 1975, \apj, 200, 402
\bibitem[Cardelli \& Meyer(1997)]{car97} Cardelli, J. A., \& Meyer, D. M.
    1997, \apjl, 477, L57
\bibitem[Cardelli et al.(1996)]{car96} Cardelli, J. A., Meyer, D. M., Jura, M.,
    \& Savage, B. D. 1996, \apj, 467, 334
\bibitem[Cardelli et al.(1991)]{car91} Cardelli, J. A., Savage, B. D, \&
    Ebbets, D. C. 1991, \apjl, 383, L23
\bibitem[Cartledge et al.(2003a)]{car3a} Cartledge, S. I. B., Lauroesch, J. T.,
    \& Meyer, D. M. 2003a, in preparation
\bibitem[Cartledge et al.(2001)]{car01} Cartledge, S. I. B., Meyer, D. M.,
    Lauroesch, J. T., \& Sofia, U. J. 2001, \apj, 562, 394
\bibitem[Cartledge et al.(2003b)]{car3b} Cartledge, S. I. B., Meyer, D. M., \&
    Lauroesch, J. T. 2003b, in preparation
\bibitem[Chan et al.(1992)]{cha92} Chan, W. F., Cooper, G., Guo, X., Burton,
    G. R., \& Brion, C. E. 1992, \pra, 46, 149
\bibitem[de Avillez \& Mac Low(2002)]{dea02} de Avillez, M. A., \& Mac Low,
    M.-M. 2002, \apj, 581, 1047
\bibitem[Diplas \& Savage(1994)]{dip94} Diplas, A., \& Savage, B. D. 1994,
    \apjs, 93, 211
\bibitem[Feltzing, Holmberg, \& Hurley(2001)]{fel01} Feltzing, S., Holmberg,
    J., \& Hurley, J. R. 2001, \aap, 377, 911
\bibitem[Grevesse \& Noels(1996)]{gre96} Grevesse, N., \& Noels, A. 1996, in
    ASP Conf. Proc. 99, Cosmic Abundances, ed. S. S. Holt \& G. Sonneborn (San
    Francisco: ASP), 116
\bibitem[Griffen \& Hutchenson(1969)]{gri69} Griffen, P. M., \& Hutchenson,
    J. W. 1969, J. Opt. Soc. Am., 59, 1607
\bibitem[Henry \& Worthey(1999)]{hen99} Henry, R. B. C., \& Worthey, G. 1999,
    \pasp, 111, 919
\bibitem[Holweger(2000)]{hol00} Holweger, H. 2001, in AIP Conf. Proc. 598,
    Solar and Galactic Composition, ed. R. F. Wimmer-Schweingruber (Berlin:
    Springer), 23
\bibitem[Howk \& Sembach(2000)]{how00} Howk, J. C., \& Sembach, K. R. 2000,
    \aj, 119, 2481
\bibitem[Johnson(1972)]{joh72} Johnson, C. E. 1972, \pra, 5, 2688
\bibitem[Mar \& Bailey(1995)]{mar95} Mar, D. P., \& Bailey, G. 1995, Proc.
    Astron. Soc. Australia, 12, 239
\bibitem[Mason(1990)]{mas90} Mason, N. J. 1990, Meas. Sci. Technol., 1, 596
\bibitem[Meyer et al.(1997)]{mey97} Meyer, D. M., Cardelli, J. A., \& Sofia, U.
    J. 1997, \apj, 490, L103
\bibitem[Meyer et al.(1998)]{mey98} Meyer, D. M., Jura, M., \& Cardelli, J. A.
    1998, \apj, 493, 222
\bibitem[Meyer et al.(1994)]{mey94} Meyer, D. M., Jura, M., Hawkins, I., \&
    Cardelli, J. A. 1994, \apj, 437, L59
\bibitem[Nowak et al.(1978)]{now78} Nowak, G., Borst, W. L., \& Fricke, J. 1978,
    \pra, 17, 1921
\bibitem[Raboud et al.(1997)]{rab97} Raboud, D., Cramer, N., \& Bernasconi, P.
    A. 1997, \aap, 325, 167
\bibitem[Raiteri et al.(1993)]{rai93} Raiteri, C. M., Gallino, R., Busso, M.,
    Neuberger, D., \& K\"appeler, F.  1993, \apj, 419, 207
\bibitem[Rolleston et al.(2000)]{rol00} Rolleston, W. R. J., Smartt, S. J.,
    Dufton, P. L., \& Ryans, R. S. I. 2000, \aap, 363, 537
\bibitem[Roy \& Kunth(1995)]{roy95} Roy, J. -R., \& Kunth, D. 1995, \aap, 294,
    432
\bibitem[Savage \& Sembach(1991)]{sav91} Savage, B. D., \& Sembach, K. R. 
    1991, \apj, 379, 245
\bibitem[Sembach \& Savage(1994)]{sem94} Sembach, K. R., \& Savage, B. D. 1994,
    \apj, 431, 201
\bibitem[Sembach \& Savage(1996)]{sem96} Sembach, K. R., \& Savage, B. D. 1996,
    \apj, 457, 211
\bibitem[Sofia et al.(1997)]{sof97} Sofia, U. J., Cardelli, J. A., Guerin, K.
    P., \& Meyer, D. M. 1997, \apj, 482, L105
\bibitem[Sofia \& Meyer(2001)]{sof01} Sofia, U. J., \& Meyer, D. M. 2001, \apj,
    554, L221
\bibitem[Sofia, Meyer, \& Cardelli(1999)]{sof99} Sofia, U. J., Meyer, D. M., \&
    Cardelli, J. A. 1999, \apj, 522, L137
\bibitem[Smartt et al.(2001)]{sma01} Smartt, S. J., Venn, K. A., Dufton, P. L.,
    Lennon, D. J., Rolleston, W. R.  J., \& Keenan, F. P. 2001, \aap, 367, 86
\bibitem[Vall\'ee(2002)]{val02} Vall\'ee, J. P. 2002, \apj, 566, 261
\bibitem[Wells \& Zipf(1974)]{wel74} Wells, W. C., \& Zipf, E. C. 1974, \pra,
    9, 568
\bibitem[Welty, Hobbs, \& York(1991)]{wel91} Welty, D. E., Hobbs, L. M., \&
    York, D. G. 1991, \apjs, 75, 425
\bibitem[Welty et al.(1999)]{wel99} Welty, D. E., Hobbs, L. M.,
    Lauroesch, J. T., Morton, D. C., Spitzer, L., \& York, D. G. 1999, \apjs,
    124, 465
\bibitem[Zeippen, Seaton, \& Morton(1977)]{zei77} Zeippen, C. J., Seaton, M. J.,
    \& Morton, D. C. 1977, \mnras, 181, 527
\end{thebibliography}
\end{document}